\documentclass[lettersize,journal]{IEEEtran}
\usepackage{amsmath,amsfonts}
\usepackage{algorithmic}
\usepackage{algorithm}
\usepackage{array}
\usepackage[caption=false,font=normalsize,labelfont=sf,textfont=sf]{subfig}
\usepackage{textcomp}
\usepackage{stfloats}
\usepackage{url}
\usepackage{verbatim}
\usepackage{graphicx}
\usepackage{fancyvrb}
\usepackage{fancyhdr}
\usepackage{cite}
\begin{document}

\title{Weight Mapping Properties of a Dual Tree Single Clock Adiabatic Capacitive Neuron}

\author{Mike Smart,  Sachin Maheshwari,~\IEEEmembership{Member,~IEEE,}  Himadri Singh Raghav,~\IEEEmembership{Member,~IEEE,} \\Alexander Serb,~\IEEEmembership{Senior Member,~IEEE} 
\thanks{This work has been, in part, funded by Defence Science and Technology Laboratory (Dstl), UK.}
\thanks{M. Smart, S. Maheshwari, H. S. Raghav, A. Serb, are with the Centre for Electronics Frontiers, School of Engineering, University of Edinburgh, Scotland, EH9 3FB, UK. (E-mail: \{msmart2, maheshwari.sachin, hraghav, aserb\}@ed.ac.uk.}
}

\IEEEpubid{
\begin{minipage}{\textwidth}
\centering \footnotesize
\vspace{40pt}This work has been submitted to the IEEE for possible publication.\\
Copyright may be transferred without notice, after which this version may no longer be accessible.
\end{minipage}
}



\maketitle

\begin{abstract}
Dual Tree Single Clock (DTSC) Adiabatic Capacitive Neuron (ACN) circuits offer the potential for highly energy-efficient Artificial Neural Network (ANN) computation in full custom analog IC designs.
The efficient mapping of Artificial Neuron (AN) abstract weights, extracted from the software-trained ANNs, onto physical ACN capacitance values has, however, yet to be fully researched. In this paper, we explore the unexpected hidden complexities, challenges and properties of the mapping, as well as, the ramifications for IC designers in terms accuracy, design and implementation. We propose an optimal, AN to ACN methodology, that promotes smaller chip sizes and improved overall classification accuracy, necessary for successful practical deployment.
Using TensorFlow and Larq software frameworks, we train three different ANN networks and map
their weights into the energy-efficient DTSC ACN capacitance value domain 
to demonstrate 100\% functional equivalency. Finally, we delve into the impact of weight quantization on ACN performance using novel metrics related to practical IC considerations, such as IC floor space and comparator decision-making efficacy.
\end{abstract}

\begin{IEEEkeywords}
Adiabatic Logic, Binary Neural Networks, Switched Capacitor, Analog Computation, Energy-Efficiency.
\end{IEEEkeywords}

\section{Introduction}
\IEEEPARstart{A}{diabatic Logic} (AL) is a low-power ASIC design technique where charge used for computation is periodically returned to a slowly alternating AC power supply~\cite{Teichmann2012,Athas1994}. This process allows for energy-efficient analog computation, compared to \mbox{non-adiabatic}, CMOS digital solutions that use fixed DC supplies. AL requires a compatible supply source for charge recovery, such as those designed using an inductor-based oscillator~\cite{Koller1992, Maksimovic2000, Maheshwari2022} or a capacitive-based step-charging circuit~\cite{Nakata2012, Raghav2016,Raghav2017}.
AL has been applied to the design of in-memory computation~\cite{Thapliyal2020}, content addressable memories~\cite{chang2011}, power drivers~\cite{Lal2000}, secure hardware design~\cite{Raghav2019}, error detection in data transmission~\cite{Maheshwari2018} and spiking networks
~\cite{Massarotto2023,Massarotto2024}.

In parallel to advances in AL, ANNs have been implemented in analog hardware using Switched Capacitor (SC) networks~\cite{Maundy1991,Bankman2017, Giacomo2023,Hajtas2000,Zhu2023}.
Early work in non-adiabatic SC neural networks, such as by Tsividis \emph{et al}~\cite{Tsividis1987}, used a proportional mapping scheme to convert AN weights to SC capacitance values. Other authors have used \mbox{non-adiabatic} charge-to-weight-to-capacitance value relationships~\cite{Maundy1991}, switch capacitance to amplifier feedback capacitance ratios~\cite{Ronchi2025}, binary weighted mappings~\cite{Valavi2019,Bankman2017} and, most recently, differential capacitance values~\cite{Massarotto2024}.
The Adiabatic Capacitive Artificial Neuron (ACAN) combined AL techniques with an SC circuit to minimize use of energy dissipating resistive elements and perform low-power, AN, dot-product computation~\cite{Maheshwari2022,Maheshwari2021}. ACAN was restricted to ANNs with positive-valued weights and a binary activation function. Similar to ACAN, the adiabatic DTSC ACN was subsequently introduced with support for both
positive and negative weights using only a minimal number of switched capacitors~\cite{Maheshwari2025}.

\begin{figure}[!t]
\centering
\includegraphics[width=0.4\textwidth]{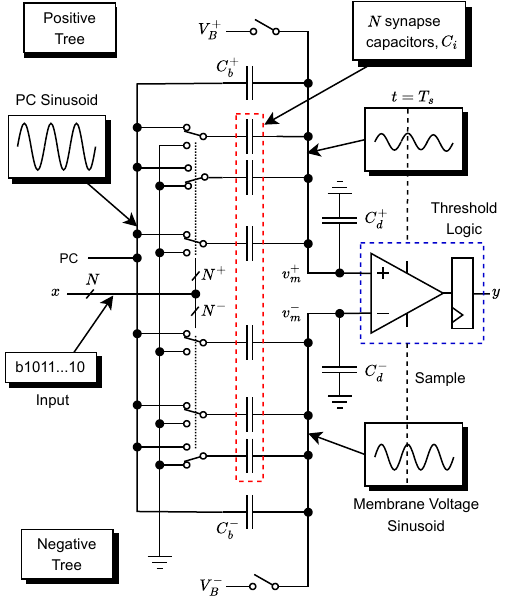}
\caption{Double Tree Single Clock (DTSC) $N$-bit differential ACN~\cite{Maheshwari2025} with binary input vector, \textbf{x}, two capacitive trees with a total of $N$ switched synapse capacitors and a single sinusoidal Power Clock (PC) supply. A Threshold Logic (TL) unit generates a binary output, $y$, by sampling, and comparing, the two membrane voltages, $v_{m}^{\pm}$, at a time, $T_{s}$.}
\label{fig:dtsc-circuit}
\end{figure}

Away from the developments in SC and AL analog hardware, there has also been significant recent research into Quantized Neural Networks (QNNs) ~\cite{Hubara2018} and Binary Neural Networks (BNNs) that use ANs with a binary activation function~\cite{Qin2020,Yuan2023}. Generally, work in this area has been driven by the desire for fast, efficient and low power digital AN hardware. Historically, training BNNs was problematic due to the requirement to back-propagate error gradients through neurons with binary outputs. The introduction of the Straight Through Estimator (STE) and STE variants~\cite{Bengio2013,Hubara2018}, however, has meant that training BNNs to perform real-world tasks, such as image classification, is now possible with open-source BNN software frameworks, such as TensorFlow-based Larq~\cite{Geiger2020}. Researchers have also shown the ability of BNNs, with suitable resources, to achieve classification rates comparable to standard ReLU-based  ANNs on well-known datasets, such as MNIST and CIFAR-10~\cite{Hubara2018}. This has opened the door for the mapping of real-world, software-trained ANs directly onto AL-based SC hardware, like the DTSC ACN.

At first glance, mapping weights from these real-world ANNs onto ACN capacitance values, using a simple proportional scheme, as before, might seem trivial. In this paper we show that, actually, a more elaborate approach is required. The paper introduces an optimal mapping that provides exact functional equivalence, whilst using a minimal number of switched capacitors. Further, we show that ANN training, quantization and mapping can have significant impact on the ultimate hardware performance, such as IC classification accuracy and chip size. The paper also introduces a range of mapping-related tools and properties to aid the IC designer.

The paper begins in Section~\ref{sec:acn_operation} with a brief ACN operational review.
Section~\ref{sec:mapping} 
then considers various approaches to mapping AN weights and details the mathematical properties and significant benefits of the preferred conditional mapping algorithm. Section~\ref{sec:ic_considerations} discusses the practical ASIC design implications of the conditional mapping, which have not been previously explored in the literature. In Section~\ref{sec:software} the paper concludes with an ACN mapping demonstration using various MNIST-trained ANNs, BNNs and QNNs.

\section{ACN operation}\label{sec:acn_operation}
To understand how an abstract artificial neuron can efficiently map onto adiabatic hardware first requires an  examination of the operation of an ACN. The DTSC ACN circuit~\cite{Maheshwari2025}, reproduced in Fig.~\ref{fig:dtsc-circuit}, uses a sinusoidal voltage supply, referred to as a Power Clock (PC), to charge two independent banks, or \emph{trees}, of \emph{synapse} capacitors on the upswing of the PC. Two sinusoidal \emph{membrane} voltages, $v_{m}^{\pm}$, one for each tree, appear on the input terminals of a Threshold Logic (TL) unit. The TL then samples and compares both membrane voltages at a time, $t=T_{s}$, typically when the sinusoidal PC voltage, $V_{pc}(t)$, reaches its peak voltage, $V_{max}$. The binary output of the TL, $y$, is ideally set to binary '1' if $v_{m}^{+} \ge v_{m}^{-}$, else it is set to binary '0'. On the PC downswing, the charge used for computation is returned to the supply. The DTSC ACN charge recovery principle remains the same as with previous ACAN work. A more detailed description of the cyclical operation can be found in ~\cite{Maheshwari2022,Maheshwari2025}.

The magnitudes of $v_{m}^{\pm}$ are determined by which synapse capacitors are allowed to contribute in each tree. This is controlled by a set of $N$ Single Pole Double Throw (SPDT) switches, shown to the left of the synapse capacitors in Fig.~\ref{fig:dtsc-circuit}. Each SPDT switch is controlled by a single bit from an \mbox{$N$-bit} binary input signal, $x$. If the $ith$ input bit, $x_{i}$, is binary '1' then the PC is allowed to propagate to the associated synapse capacitor $C_{i}$ and contribute to the membrane voltage. Otherwise the synapse capacitor is connected to ground.
More formally, the $N$ synapse capacitors are split into two disjoint sets of $N^{+}$ capacitance values, $C^{+}_{i}$ where $i\in I^{+}$, and $N^{-}$ capacitance values, $C^{-}_{i}$ where $i\in I^{-}$. $I^{\pm}$ are disjoint subsets of the indexing set $I=\{0..,N-1\}$, such that $N = N^{+} + N^{-}$.

In addition to the synapse capacitors, there are two extra capacitors on each tree: a bias capacitor, $C_{b}$, and a ballast capacitor, $C_{d}$. The bias capacitors always allow the PC to propagate, providing a minimal swing of the membrane voltage, even when all inputs are zero. The ballast capacitors can be used to control the maximum swing of $v_{m}^{\pm}$.

Finally, each capacitive tree has a DC bias voltage, $V_{B}$. Before computation the ACN is put into an initial reset state by closing the two Transmission Gate (TG) switches connected to $V_{B}^{\pm}$. This initializes both membrane voltages to a known fixed value. Once the reset is complete, these TG switches are re-opened to allow for computation.

Based on the previous definitions, and using standard capacitive voltage division, it is possible to formulate an idealized equation for  $v_{m}^{\pm}$, at a time $t$, on each tree as follows
\begin{equation}
\label{eq:vm_general_equation}
v_{m}^{\pm}(t) = V_{B}^{\pm} + V_{pc}(t) \sum_{i \in I^{\pm}} \frac{C_{i}^{\pm}x_{i}+C_{b}^{\pm}}{C_{T}^{\pm}+C_{b}^{\pm}+C_{d}^{\pm}}
\end{equation}
where $C_{T}^{\pm}$ represents the summed, parallel, synapse capacitance values in each tree.
The membrane voltages in \mbox{(\ref{eq:vm_general_equation})} can be alternatively expressed more simply as
\begin{equation}
\label{eq:vm_general_equation_using_con}
v_{m}^{\pm}(t) = V_{B}^{\pm} + V_{pc}(t) \frac{C_{on}^{\pm}}{C_{on}^{\pm}+C_{off}^{\pm}}
\end{equation}
where $C^{\pm}_{on}$ is the sum of the switched on capacitance values ($x_{i}=1$) plus the bias capacitance in each tree, and $C^{\pm}_{off}$ is the sum of all the switched off capacitance values ($x_{i}=0$) connected to ground, plus the ballast capacitance, $C^{\pm}_{d}$.

Several differences exist between the previous ACAN design and the DTSC ACN. There are now two capacitor trees along with bias capacitors. Also, 
there is no TL absolute DC reference voltage and the Single Pole Single Throw (SPST) synapse switches have been replaced by SPDT switches. Importantly, the number of synapse capacitors remains as $N$ in DTSC compared to other SC signed-weight solutions using $2N$ differential capacitance values~\cite{Massarotto2024}. 

\section{AN To ACN Mapping}\label{sec:mapping}
In order to replicate the mathematical operation of a software-trained AN, the weights and biases need to be mapped to the capacitance values in the DTSC ACN circuit. The mapping requires that the same binary output, $y$, from the software AN and the hardware ACN, is generated for the same binary input vector, $\textbf{x}$, for all possible valid input sequences.

For AL charge recovery to work, a high impedance, binarizing comparator is used in the ACN TL. This architecture is inherently well-suited for mapping ANs that utilize binary activation functions. A compatible AN accepts $\textbf{x}$ as input and generates a single bit output, $y$, as such
\begin{equation}
\label{eq:sw_equation}
y = \begin{cases}
    1, & \text{if } \textbf{w}\cdot\textbf{x} =  \sum_{i \in I} w_{i}x_{i} \ge \tau\\
    0, & \text{otherwise}
\end{cases}
\end{equation}
\noindent where \(w_{i}\) are $N$ synapse weights that form a weight vector \(\mathbf{w}\) and \(\tau\) is a bias value. The weights and bias may be real-valued ($w_{i},\tau \in \mathbb{R}$), quantized or binary in nature.

The $N$ weights can be split into two sets of $N^{+}$ positive-valued weights, $w^{+}_{i}$, and $N^{-}$ negative-valued weights, $w^{-}_{i}$.
To perform the mapping, the positive-valued $w^{+}_{i}$, need to be converted to a set of $N^{+}$ positive-valued ACN capacitance values, $C_{i}^{+}$ in the positive tree of the DTSC ACN. Similarly, the negative-valued $w^{-}_{i}$, need to be mapped to $N^{-}$ \emph{positive}-valued capacitance values, $C_{i}^{-}$, in the negative tree.

\subsection{Conditional mapping}
\label{sec:conditional_mapping}
To map software AN functionality to an ACN, the dot-product \emph{condition} defined in \mbox{(\ref{eq:sw_equation})} must be preserved. As such, a \textbf{\emph{conditional mapping}} approach is proposed that solely focuses on
whether the ACN will output a binary '1' when the relative condition $v_{m}^{+} \ge v_{m}^{-}$ is met, rather than absolute voltage levels. Assuming $V_{B}^{+} = V_{B}^{-}$ then \mbox{(\ref{eq:vm_general_equation})} can be expressed as
\begin{equation}
\label{eq:dtsc_proof_dtsc_hw_equation}
V_{pc}(t) \left[ \sum_{{i \in I^{+}}} \frac{C_{i}^{+}x_{i}+C_{b}^{+}}{C_{A}^{+}} \right] \ge V_{pc}(t) \left[ \sum_{i \in I^{-}} \frac{C_{i}^{-}x_{i}+C_{b}^{-}}{C_{A}^{-}} \right]
\end{equation}
where $C_{A}^{\pm} = C_{T}^{\pm}+C_{b}^{\pm}+C_{d}^{\pm}$. The ACN will output binary '0' when this condition is not met. The benefit of SPDT switches in ACN means that both $C_{A}^{\pm}$ terms remain constant, regardless of input. This, significantly, provides an input-output linearity, not present in ACAN, such that  (\ref{eq:dtsc_proof_dtsc_hw_equation}) can be refactored by multiplying both sides by $C_{A}^{-}/C_{T}^{-}$, collecting bias terms, multiplying the LHS by  $C_{T}^{+}/C_{T}^{+}$ and canceling $V_{pc}(t)$
\begin{equation}
\label{eq:dtsc_proof_hw_refactor_2}\frac{C_{T}^{+}C_{A}^{-}}{C_{T}^{-}C_{A}^{+}} \sum_{i \in I^{+}} \frac{C_{i}^{+}x_{i}}{C_{T}^{+}}  
\ge \frac{C_b^{-}}{C_{T}^{-}} - \frac{C_{b}^{+}C_{A}^{-}}{C_{T}^{-}C_{A}^{+}} + \sum_{i \in I^{-}} \frac{C_{i}^{-}x_{i}}{C_{T}^{-}}
\end{equation}

Returning to the AN model, condition (\ref{eq:sw_equation}) can be split into two positive and negative components and rewritten as
\begin{equation}
\label{eq:reworked-sw-model}
\sum_{i \in I^{+}} w_{i}^{+}x_{i} \ge \tau + \sum_{i \in I^{-}} |w_{i}^{-}|x_{i}
\end{equation}
Then by replacing the now positive-valued weighting terms on both sides of  condition (\ref{eq:reworked-sw-model}) with
\begin{equation}
\label{eq:general_syn_weight_map}
|w_{i}^{\pm}| = C_{i}^{\pm}w_{T}^{\pm}/C_{T}^{\pm}
\end{equation}
where \(w_{T}^{\pm}\) is a positive-valued weight vector scaling factor
\begin{equation}
w_{T}^{\pm} = \sum_{i \in I} |w_{i}^{\pm}|
\end{equation}
a software-based AN mapping condition is derived, which can be mapped directly onto the hardware condition  (\ref{eq:dtsc_proof_hw_refactor_2}).

\begin{equation}
\label{eq:dtsc_proof_map_equation_2}
\frac{w_{T}^{+}}{w_{T}^{-}}  \sum_{i \in I^{+}} \frac{C_{i}^{+}x_{i}}{C_{T}^{+}} \ge \frac{\tau}{w_{T}^{-}} + \sum_{i \in I^{-}} \frac{C_{i}^{-}x_{i}}{C_{T}^{-}}
\end{equation}
Comparing the AN condition (\ref{eq:dtsc_proof_map_equation_2}) with ACN condition (\ref{eq:dtsc_proof_hw_refactor_2}):
\begin{equation}
\label{eq:dtsc_proof_map_scaling_term}
\frac{C_{T}^{+}C_{A}^{-}}{C_{T}^{-}C_{A}^{+}} 
 = \frac{w_{T}^{+}}{w_{T}^{-}}
\end{equation}
\begin{equation}
\label{eq:dtsc_proof_map_bias_term}
\frac{C_{b}^{-}}{C_{T}^{-}} - \frac{C_{b}^{+}C_{A}^{-}}{C_{T}^{-}C_{A}^{+}} = 
\frac{\tau}{w_{T}^{-}}
\end{equation}

Under the condition $C_{T}^{+}/C_{T}^{-}=w_{T}^{+}/w_{T}^{-}=C_{T}/w_{T}$, where $w_{T}=w_{T}^{+}+w_T^{-}$ and $C_{T}=C_{T}^{+}+C_{T}^{-}$, then \mbox{(\ref{eq:dtsc_proof_map_scaling_term})} simplifies to the balanced capacitive tree condition $C_{A}^{-}=C_{A}^{+}$. The synapse capacitance values can then be determined from \mbox{(\ref{eq:general_syn_weight_map})} as
\begin{equation}
\label{eq:simple-prop-mapping}
C_{i}^{\pm} =  C_{T}|w_{i}^{\pm}
|/w_{T}
\end{equation}
and \mbox{(\ref{eq:dtsc_proof_map_scaling_term})}
can be expanded and rewritten in terms of $C_{d}^{-}$ as
\begin{equation}
C_{d}^{-} = \frac{(w_{T}^{+}-w_{T}^{-})C_{T}}{w_{T}} + C_{b}^{+} + C_{d}^{+} - C_{b}^{-}
\end{equation}
which holds for $C_{d}^{-} \ge 0$ when $w_{T}^{+} \ge w_{T}^{-}$ and
\begin{equation}
\label{eq:alg_cond_mapping_1}
C_{d}^{-} = \frac{(w_{T}^{+}-w_{T}^{-})C_{T}}{w_{T}} + C_{b}^{+} \text{  and  } C_{d}^{+} = C_{b}^{-}
\end{equation}

Rewriting (\ref{eq:dtsc_proof_map_scaling_term}) in terms of $C_{d}^{+}$, with the same conditions, then 
$C_{d}^{+} \ge 0$ holds when $w_{T}^{-} \ge w_{T}^{+}$ and
\begin{equation}
\label{eq:alg_cond_mapping_2}
C_{d}^{+} = \frac{(w_{T}^{-}-w_{T}^{+})C_{T}}{w_{T}} + C_{b}^{-}  \text{   and   }  C_{d}^{-} = C_{b}^{+}
\end{equation}

From \mbox{(\ref{eq:dtsc_proof_map_bias_term})} the condition $C_{b}^{-} \ge 0$ holds for $\tau \ge 0$ when
\begin{equation}
C_{b}^{+} = 0 \text{ and } C_{b}^{-} = \tau C_{T}/w_{T}
\end{equation}
as $\tau$, $C_{T}$ and $w_{T}^{\pm}$ are all positive-valued by definition. Similarly, by using the $C_{A}$ ratio in \mbox{(\ref{eq:dtsc_proof_map_scaling_term})},
the condition $C_{b}^{+} \ge 0$ holds for $\tau < 0$ when
\begin{equation}
C_{b}^{-} = 0 \text{ and } C_{b}^{+} = \lvert \tau \lvert C_{T}/w_{T}
\end{equation}

We now have the foundations necessary to calculate all the physical ACN capacitance values  based on AN model parameters. Under ideal conditions the ACN will produce an identical output as the AN, given the same binary input vector, $\textbf{x}$.
The value of $C_{T}$ can be considered as a design choice, and a
\emph{scaling constant} of the physical ACN.

The proposed conditional mapping scheme has demonstrated that the process is non-trivial, requiring specific values of $C_{d}^{\pm}$ for a robust mapping. The mapping also raises a number of important points that we will explore, in detail, in the remainder of this paper:
1) the conditional mapping ensures AN and ACN output equivalency but does not define the relationship between the ACN $v_{m}^{\pm}$ voltages and the originating AN $\textbf{w}\cdot\textbf{x}$ dot product;
2) there is no indication whether the mapping is in some sense optimal e.g. in terms of minimal $C_{d}$;
3) from a mapping perspective, the bias capacitors could be treated as always-on capacitors;
4) a new design choice, $C_{T}$, has been introduced but with no guidance on what value a designer should use;
and lastly, 5)  
although $C_{A}^{\pm}$ in 
\mbox{(\ref{eq:dtsc_proof_dtsc_hw_equation})}
is constant due to the choice of SPDT synapse switches and provides linearity with respect to the input, $x$, it remains a function of $C_{i}$ and, as such, places a normalizing constraint on the computation not present in the software AN.

\subsection{Vector space}
\begin{figure}[ht]
  \centering
  \includegraphics[width=0.4\textwidth]{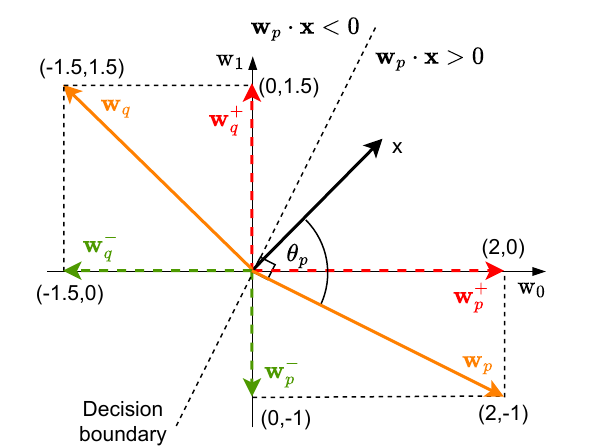}
    \caption{$N=2$ vector space with input vector, $\mathbf{x} = (1,1)$ and two example weight vectors ($\mathbf{w_{p}}$ and $\mathbf{w_{q}}$) that occupy  $\mathbb{R}^{2}$. Each weight vector (orange) represents a different AN, each of which can be resolved into two vectors: a vector with positive-valued components (red) and a vector with all negative-valued components (green).}
    \label{fig:dtsc_mapping_vector_simple}
\end{figure}

To better understand the points discussed in \ref{sec:conditional_mapping} this section examines the AN to DTSC ACN conditional mapping in $\mathbb{R}^{N}$ vector weight space. To aid visualization, a simple $N=2$ vector space with $\tau=0$ is considered, as demonstrated in Fig.~\ref{fig:dtsc_mapping_vector_simple}.
In this case, the AN output, $y$, as specified by \mbox{(\ref{eq:sw_equation})}, is '1' when \mbox{$\textbf{w}\cdot\textbf{x} = \lvert \textbf{w} \lvert \lvert \textbf{x} \lvert \text{cos} \, \theta \ge 0$}. This occurs when the angle, $\theta$, between  $\textbf{w}$ and $\textbf{x}$ satisfies $-90^{\circ} \le \theta \le 90^{\circ}$. So, in this AN implementation the weight vector magnitude, $\lvert \textbf{w} \lvert$, is  irrelevant to the output, $y$, and only $\theta$ needs to be considered. Consequently, only the direction of $\textbf{w}$ needs to be preserved in the AN to ACN mapping to generate an equivalent $y$.

To match the dual tree structure of the target DTSC ACN architecture  requires that the weight vector, $\textbf{w}$, be resolved into two orthogonal $\mathbb{R}^{N}$ vectors, $\textbf{w}^{+}$ and $\textbf{w}^{-}$, representing all the positive-valued and all the negative-valued weights respectively, such that $\textbf{w} = \textbf{w}^{+} + \textbf{w}^{-}$ and $\textbf{w}^{+} \cdot \textbf{w}^{-} = 0$. 
More formally, $\textbf{w}^{+} = (w_{0}^{+},...w_{i}^{+}..,w_{N-1}^{+})$ where $w_{i}^{+} = w_{i}$ if $w_{i} > 0$ else $0$, and conversely, $w_{i} < 0$ for $\textbf{w}^{-}$.

\begin{figure}[ht]
  \centering
  \includegraphics[width=0.45\textwidth]{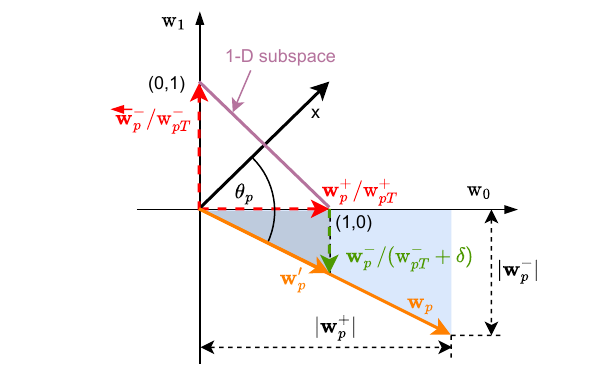}
    \caption{Example weight vector $\mathbf{w_{p}}$ mapped onto a DTSC compatible weight space where the normalized weight vectors are constrained to a positive-valued $N-1$ dimensional subspace (purple). A reconstructed vector, $\mathbf{w'_{p}}$, can be made to point in same direction ($\theta_{p}$) as the original $\mathbf{w_{p}}$ through vector subtraction and use of a correction term, $\delta$. Note that $|\mathbf{w'_{p}}| \neq |\mathbf{w_{p}}|$.}
    \label{fig:dtsc_mapping_vector_mapped}
\end{figure}

Each of these  $\textbf{w}^{\pm}$ vectors now needs to be normalized by $w_{T}^{\pm}$. This is required to match the normalizing constraint generated by ACN capacitive voltage division in each DTSC tree. Next, as the negative-valued weights are required to map to positive-valued physical capacitance values,  $\textbf{w}^{-}$ is flipped to generate a new vector, $\overleftarrow{\textbf{w}}^{-} = -1 \textbf{w}^{-}$. Together, normalization and flipping, constrains the two resulting vectors to an $N-1$ dimensional subspace defined by a $\mathbb{R}_{\ge 0}^{N}$ hyperplane, as demonstrated in Fig.~\ref{fig:dtsc_mapping_vector_mapped}. This hyperplane intersects all N unit vectors in the weight space. This means both of these modified vectors are now in the positive-only quadrant, such that $\textbf{w}^{+}/w_{T}^{+} \in \mathbb{R}_{\ge 0}^{N}$ and $\overleftarrow{\textbf{w}}^{-}/w_{T}^{-} \in \mathbb{R}_{\ge 0}^{N}$. Orthonormality of the modified vectors seen in Fig.~\ref{fig:dtsc_mapping_vector_mapped} is unlikely for $N>2$ due to the often non-sparse nature of trained weight vectors.

Although now in a suitable form for DTSC mapping, the resulting cumulative  vector $ \textbf{w}^{+}/w_{T}^{+} - \overleftarrow{\textbf{w}}^{-}/w_{T}^{-}$ does not point in the same direction as $\textbf{w}$, due to the differing normalization scalars, $w_{T}^{\pm}$, typical in trained ANs. To correct the direction a positive-valued scalar value, $\delta$, can be added to one of the normalizing terms. This scaling preserves a weight form that can still map to the two DTSC voltage division trees. In the  Fig.~\ref{fig:dtsc_mapping_vector_mapped} example, as $w_{T}^{+} > w_{T}^{-}$, the negative term is appropriate as $\delta$ can only scale down the vector magnitude. This correction term means, with a specific value of $\delta$, that $\textbf{w}' = \textbf{w}^{+}/w_{T}^{+} - \overleftarrow{\textbf{w}}^{-}/(w_{T}^{-}+\delta)$ will point in the same direction as $\textbf{w}$.
From the proportional scaling properties of the two similar triangles shown in blue in Fig.~\ref{fig:dtsc_mapping_vector_mapped}, it is known that the ratio of the two triangles lengths are equivalent. Specifically, $\lvert w^{+}\lvert/w_{T}^{+}\lvert w^{+}\lvert =\lvert w^{-}\lvert/(w_{T}^{-}+\delta)\lvert w^{-}\lvert$ and, as such, $\delta = w_{T}^{+} - w_{T}^{-}$. As will be shown later, this also holds when $N > 2$.

Returning to the ACN, the $N$ capacitance values, $C_{i}$, can hypothetically be expressed as a capacitive vector, $\textbf{c}$, with signed values. Using proportional weight scaling then  \mbox{$\textbf{c}= \alpha\textbf{w}$}, 
where $\alpha$ is a positive-valued scalar and, as such, \mbox{$\textbf{c}^{\pm}= \alpha\textbf{w}^{\pm}$} with $C_{T}^{\pm}=\alpha w_{T}^{\pm}$.
As $\tau = 0$ we let $C_{b}^{\pm}=0$ and assuming $C_{d}^{+} = 0$ the ACN \emph{differential} membrane voltage $\Delta v_{m} = v_{m}^{+} - v_{m}^{-}$ when $V_{pc}(t)=V_{max}$ can be expressed as 
\begin{equation}
\label{eq:dtsc_differ_voltage}
\Delta v_{m} = V_{max} \left[ \sum_{{i \in I^{+}}} \frac{C_{i}^{+}x_{i}}{C_{T}^{+}} - \sum_{i \in I^{-}} \frac{C_{i}^{-}x_{i}}{C_{T}^{-}+C_{d}^{-}} \right]
\end{equation}
or, in vector notation with the input resolved as $\textbf{x} = \textbf{x}^{+} + \textbf{x}^{-}$
\begin{equation}
\label{eq:dtsc_differ_voltage_vector}
\Delta v_{m} = V_{max} \left[
\textbf{c}^{+}\cdot\textbf{x}^{+}/C_{T}^{+} - \overleftarrow{\textbf{c}}^{-}\cdot\textbf{x}^{-}/(C_{T}^{-}+C_{d}^{-})\right]
\end{equation}
or by substitution and given that $\textbf{c}^{+}\cdot\textbf{x}^{-} = \overleftarrow{\textbf{c}}^{-}\cdot\textbf{x}^{+} = 0$
\begin{equation}
\label{eq:dtsc_differ_voltage_vector2}
\Delta v_{m} = V_{max} \left[
\textbf{c}^{+}\cdot\textbf{x}/C_{T}^{+} - \overleftarrow{\textbf{c}}^{-}\cdot\textbf{x}/(C_{T}^{-}+C_{d}^{-})\right]
\end{equation}
This means $\Delta v_{m}$ is proportional to  $\textbf{C} \cdot \textbf{x}$, where $\textbf{C}$ is a hypothetical normalized capacitance value vector defined by
\begin{equation}
\label{eq:detivation_pf_c_normaliszed}
\textbf{C} = \textbf{c}^{+} /C_{T}^{+} - \overleftarrow{\textbf{c}}^{-}/(C_{T}^{-}+C_{d}^{-})
\end{equation}
Like $\textbf{w}'$ the vector $\textbf{C}$ will preserve the direction of $\textbf{c}$, and thus $\textbf{w}$, when the correction $C_{d}^{-} = C_{T}^{+}-C_{T}^{-}$ or $C_{T}(w_{T}^{+}-w_{T}^{-})/w_{T}$, is used, as previously shown algebraically. Repeating the process when \mbox{$w_{T}^{-} > w_{T}^{+}$} is a trivial exercise.


\subsection{Capacitive dot-product}
As shown, the AN weight vector, $\textbf{w}$, can be conditionally mapped to the DTSC ACN as positive-valued capacitance values.
However, although the direction of $\textbf{w}$ and $\textbf{C}$ are equivalent, their magnitudes tend to differ and generally $\lvert \textbf{w} \lvert \neq \lvert \textbf{C} \lvert$ and, as such, $\textbf{w} \cdot \textbf{x} \neq \textbf{C} \cdot \textbf{x}$.
Continuing with $\tau=0$, the sampled TL differential input voltage is given by $V_{max}\textbf{C} \cdot \textbf{x}$ or 
\begin{equation}
\label{eq:dvm_as_expanded_dot_product}
    \Delta v_{m} = V_{max}\lvert \textbf{C} \lvert \lvert \textbf{x} \lvert \text{cos} \:\theta
\end{equation}
Thus, both the magnitude,  $\lvert \textbf{C} \lvert$, and relative direction, $\theta$, of the capacitive vector, $\textbf{C}$, may have practical implications.

\begin{figure}[ht]
  \centering
  \includegraphics[width=0.4\textwidth]{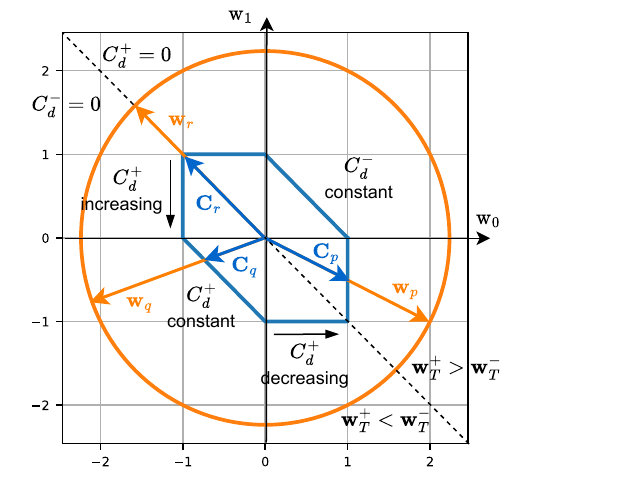}
    \caption{Three examples of $N=2$ weight vectors ($\mathbf{w_{p}},\mathbf{w_{q}},\mathbf{w_{r}}$) all with the same magnitude, $\lvert \textbf{w} \lvert = \sqrt{5}$, but differing directions, $\phi$. Also shown are the corresponding conditionally-mapped capacitance vectors ($\mathbf{C_{p}},\mathbf{C_{q}},\mathbf{C_{r}}$). The ballast capacitor sizes, $C_{d}^{\pm}$, are also shown to vary with $\phi$.}
    \label{fig:dtsc_mapping_vector_c_phi}
\end{figure}

An $N=2$ weight vector with a constant magnitude, $\lvert \textbf{w} \lvert$, and varying absolute direction, $\phi$, will trace out a circular path, as shown in 
Fig.~\ref{fig:dtsc_mapping_vector_c_phi}. Due to normalization the path of  \textbf{C} is, however, not circular. In this superficial $\mathbb{R}^{2}$ example the hexagonal path, shown in blue, for $\textbf{C}$ is traced. In these non-typical cases, where $N^{\pm}=1$, one tree will include a single capacitor with no ballast, making the weight value ineffective. Typically, the path when $N^{\pm}>2$ will form a hypercube, with  $\lvert \textbf{C} \lvert$ being a function of  $\phi$. This implies the capacitive dot-product, and consequently, $\Delta v_{m}$, will change, perhaps unexpectedly, for different ACNs, dependent on ANN training. It should be noted with conditional mapping, that the same $\textbf{C}$ path will be traced, regardless of $\lvert \textbf{w} \lvert$.

\begin{figure}[ht]
  \centering
  \includegraphics[width=0.35\textwidth]{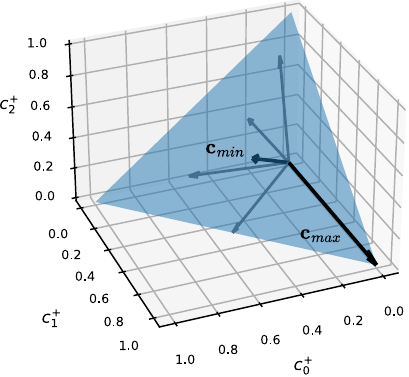}
    \caption{$N^{+}=3$ space with six example vectors constrained to triangular 2D plane (light blue). Vector $\textbf{c}_{min}$ has minimum length when $c_{0}^{+}=c_{1}^{+}=c_{2}^{+}$ and vector $\textbf{c}_{max}$ is longest when all weights are zero bar one.}
    \label{fig:dtsc_n_plus_3d}
\end{figure}
Fig.~\ref{fig:dtsc_n_plus_3d} shows a $\mathbb{R}_{>0}^{3}$ vector space with six example $\textbf{c}^{+}/C_{T}^{+}$ vectors constrained to a 2D surface. Weight vectors from trained ANs will vary in their distributions. An AN could include small numbers of large-valued weights combined with numerous smaller-valued weights. These distributions can have $\lvert \textbf{c}^{+}/C_{T}^{+} \lvert$ closer to the maximum of 1. Conversely, $\lvert \textbf{c}^{+}/C_{T}^{+} \lvert$ will be minimum when the vector is orthogonal to the hyperplane, or when $C_{i}^{+} = C_{T}^{+}/N^{+} \: \forall i \in I^{+}$ which gives a magnitude of $1/\sqrt{N^{+}}$. Thus, $\Delta v_{m}$ will become increasingly small with increasing $N^{+}$, and potentially, further reduced in the case of binarized weights.

Finally, from the $\Delta v_{m}$ definition in \mbox{(\ref{eq:dtsc_differ_voltage})} it is clear that \mbox{$-1 \le \textbf{C} \cdot \textbf{x} \le 1$}. This condition can potentially lead to saturation of $\textbf{C} \cdot \textbf{x}$ compared to the sinusoidal $\textbf{w} \cdot \textbf{x}$ as $\phi$ varies. However, this is increasingly rare as $N^{\pm}$ increases and, as it is only the sign of the dot product that impacts the binary output, can generally be ignored.

\subsection{Ballast capacitance size}

\begin{figure}[ht]
  \centering
  \includegraphics[width=0.4\textwidth]{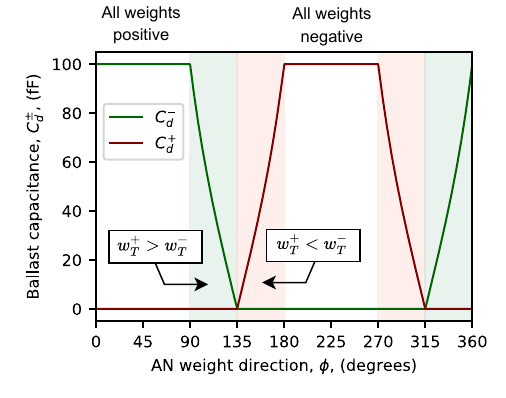}
    \caption{Variation in  $C_{d}^{\pm}$ where $N=2$ and $\tau=0$ with varying $\phi$ and $C_{T}=100fF$. Ballast size is independent of the weight magnitude, $\lvert \textbf{w} \lvert$.}
    \label{fig:dtsc_mapping_vector_cd}
\end{figure}

As seen in Fig.~\ref{fig:dtsc_mapping_vector_c_phi}, and more clearly in Fig.~\ref{fig:dtsc_mapping_vector_cd}, the size of  $C_{d}^{\pm}$ will vary with the original weight direction, $\phi$.  The conditional mapping equations \mbox{(\ref{eq:alg_cond_mapping_1})} and \mbox{(\ref{eq:alg_cond_mapping_2})} show that as $w_{T}^{\pm} \to 0$ then $C_{d}^{\pm} \to C_{T}$. In the unlikely extreme, where all weights are all positive or all negative, then $C_{d}^{\pm}$ will be zero or $C_{T}$.

\begin{figure}[ht]
  \centering
  \includegraphics[width=0.45\textwidth]{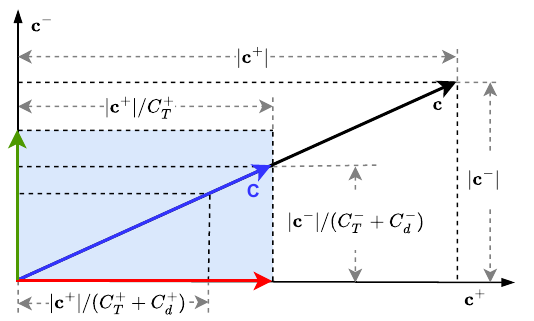}
    \caption{The $\mathbb{R}^{2}$ subspace spanned by the orthogonal $\textbf{c}^{\pm}$ vectors in $\mathbb{R}^{N}$. The hypothetical normalized capacitive vector will be constrained to the area shaded in blue. Different values of $C_{d}^{\pm}$ are capable of preserving the direction of the scaled capacitive vector, $\textbf{c}=\alpha\textbf{w}$.
    The $\tau=0$, conditionally-mapped, $\textbf{C}$ is shown with  $\max \lvert C \lvert$ and $\min \sum C_{d}^{\pm}$.
    }
    \label{fig:dtsc_mapping_vector_mapped_general}
\end{figure}

Fig.~\ref{fig:dtsc_mapping_vector_mapped_general} shows the more general case for an N-bit ACN \mbox{($\tau=0$)} where only an $\mathbb{R}^{2}$ subspace spanned by the orthogonal $\textbf{c}^{\pm}$ vectors is shown. The value of $C_{d}^{-}$ can be determined, as before, using the scaling properties of similar triangles, with
$
    \lvert c^{-}\lvert/(C_{T}^{-}+C_{d}^{-}){\lvert c^{-}\lvert} = \lvert c^{+}\lvert/C_{T}^{+}{\lvert c^{+}\lvert}
$
which gives $C_{d}^{-} = C_{T}^{+}-C_{T}^{-}$. Fig.~\ref{fig:dtsc_mapping_vector_mapped_general} shows that the conditional mapping generates the largest possible $\lvert \textbf{C} \lvert$ for a given $\textbf{w}$. Smaller $\lvert \textbf{C} \lvert$ will occur when there is ballast capacitance on both trees.

To summarize, the conditional mapping approach from the perspective of $\Delta v_{m}$ and $C_{d}$ can be considered optimal, but these values remain dependent on the original weight vector direction. It is worth a look at some alternative mappings.

\subsection{ReLU and alternate DTSC mappings}
An alternative mapping is possible by considering that (\ref{eq:dtsc_proof_map_scaling_term}) also holds for $w_{T}^{+} \ge w_{T}^{-}$ when
\begin{equation}
C_{d}^{+} = w_{T}^{-}C_{T}/w_{T} + C_{b}^{-}  \text{  and  }
C_{d}^{-} = w_{T}^{+}C_{T}/w_{T}
\end{equation}
which also works for the case where $w_{T}^{+} < w_{T}^{-}$. Using both ballast capacitors is perhaps a more \textbf{\emph{balanced mapping}}, but as indicated in Fig.~\ref{fig:dtsc_mapping_vector_mapped_general}, the total capacitance, $\sum C_{d}^{\pm}$, is much higher ($\ge C_{T}$) than the conditional mapping inferring more energy and chip size, as well as, smaller-valued $\Delta v_{m}$.

For IC designers considering physical layout or energy, it may be appropriate to differentiate between the switched $C_{i}$ and the non-switched capacitors. However, mathematically it is simpler to incorporate the bias, $\tau$, into an $N+1$ weight vector with $w_{N}=-\tau$ and $x_{N}=1$. When $w_{N+1,T}^{+} \ge w_{N+1,T}^{-}$ a \textbf{\emph{conditional vectored-bias mapping}} is defined with
\begin{equation}
C_{d}^{-} = \frac{(w_{N+1,T}^{+}-w_{N+1,T}^{-})C_{N+1,T}}{w_{N+1,T}} \text{  and  } C_{d}^{+} = 0
\end{equation}

One of the ballast capacitors is now a no-fit, which removes a capacitor and overheads. Yet, as will be seen later, there are practical reasons for instantiating both $C_{d}^{\pm}$ and to use a non-vectored $C_{b}$ mapping. Regardless, both approaches to mapping $\tau$, the bias can be included in $\textbf{C}$ and \mbox{(\ref{eq:dvm_as_expanded_dot_product})} will hold.

As seen previously,  $\textbf{C} \cdot \textbf{x} \neq \textbf{w} \cdot \textbf{x}$ with conditional mapping. This is generally fine when using a binary activation function. 
However, for completeness and comparison, it is possible to match the AN and ACN dot products, as would be required for a ReLU activation function.
Using a vectored-bias approach with $dim(\textbf{x}) = dim(\textbf{w}) =  dim(\textbf{C}) = N+1$ a \textbf{\emph{ReLU mapping}} that satisfies $\textbf{C} \cdot \textbf{x} = \textbf{w} \cdot \textbf{x}$ can be achieved as follows
\begin{equation}
    V_{max}
    \left[\frac{\textbf{c}^{+} \cdot \textbf{x}^{+}}{C_{T}^{+} + C_{d}^{+}} - 
\frac{\textbf{c}^{-} \cdot \textbf{x}^{-}}{C_{T}^{-} + C_{d}^{-}}\right] = \textbf{w} \cdot \textbf{x} =  \textbf{w}^{+} \cdot \textbf{x}^{+} - \overleftarrow{\textbf{w}}^{-} \cdot \textbf{x}^{-}
\end{equation}

Considering each term separately
\begin{equation}
    V_{max} \frac{\textbf{c}^{\pm} \cdot \textbf{x}^{\pm}}{C_{T}^{\pm} + C_{d}^{\pm}}
    = \textbf{w}^{\pm} \cdot \textbf{x}^{\pm}
\end{equation}

With linear vector scaling $\textbf{c}^{\pm} = C_{T}^{\pm}\textbf{w}^{\pm}/w_{T}^{\pm}$

\begin{equation}
\frac{V_{max}C_{T}^{\pm}}{w_{T}^{\pm}} \frac{\textbf{w}^{\pm} \cdot \textbf{x}^{\pm}}{C_{T}^{\pm} + C_{d}^{\pm}} 
    = \textbf{w}^{\pm} \cdot \textbf{x}^{\pm}
\end{equation}
\begin{equation}
    \frac{V_{max}C_{T}^{\pm}}{w_{T}^{\pm}}  
    = C_{T}^{\pm} + C_{d}^{\pm}
\end{equation}
\begin{equation}
    C_{d}^{\pm} = 
    \frac{C_{T}}{w_{T}}(V_{max} - w_{T}^{\pm})
\end{equation}

For the ballast capacitors to be realizable in the ReLU mapping the conditions $w_{T}^{\pm} \le V_{max}$ must hold.
This places a restriction on the trained weight vector. Also, in this mapping there is a dependency on the absolute voltage, $V_{max}$.

\section{Mapping Properties and IC Considerations}\label{sec:ic_considerations}
This section discusses the various mathematical properties of the mapped DTSC ACN circuit. We also discuss some of the practical considerations to an IC designer considering incorporating adiabatic DTSC ACNs into an ASIC design.

\subsection{Computational accuracy}
Mapping ACNs requires migrating numerical representations of weights in software to physical IC capacitance values. 
This infers all the standard device manufacturing inaccuracies, such as mismatch~\cite{Lin2013,Lin2013_conf,Abusleme2012}, process and thermal variations~\cite{Wakimoto2011,Wang2008}, technology constraints, noise sources, and unwanted parasitic capacitive effects, especially on the $v_{m}$ node that do not occur in software ANs.

Standard technology library, Metal-Insulator-Metal (MIM) and custom Metal-Oxide-Metal (MOM) capacitor designs in full custom ICs allow implementation of both real-valued and quantized capacitance values. Technology constraints, such as the granular geometry restrictions imposed by the chip manufacturer, can possibly introduce small quantization errors. However, careful quantization during AN weight training can mitigate for these inaccuracies to an extent.

\begin{figure}[ht]
  \centering
  \includegraphics[width=0.45\textwidth]{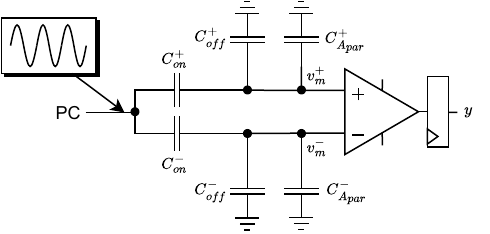}
    \caption{Simplified DTSC representation for a specific input. Parallel synapse and bias capacitors connected to the PC in each tree are replaced with a single capacitor, $C_{on}^{\pm}$, and parallel synapse and ballast capacitors connected to ground in each tree are replaced with a single capacitor, $C_{off}^{\pm}$, as defined by \mbox{(\ref{eq:vm_general_equation_using_con})}. Total parasitic capacitances on each $v_{m}$ node, $C_{A_{par}}^{\pm}$, is introduced.}
    \label{fig:dtsc_pos_neg_with_para}
\end{figure}

Parasitic capacitance on the $v_{m}$ nodes, as shown in Fig.~\ref{fig:dtsc_pos_neg_with_para}, especially when instantiating small-valued capacitors for the process node used, have the potential to significantly impact computation. The accumulated parasitic capacitances to ground, $C_{A_{par}}^{\pm}$, from all ACN capacitors connected to $v_{m}$, can be non-negligible. A solution is to use tool-generated estimates for $C_{A_{par}}^{\pm}$ and use \emph{parasitic-compensated} ballast capacitance values, as $C_{A_{par}}^{\pm}$ is in parallel with $C_{d}^{\pm}$. An issue may occur if $C_{A_{par}}^{\pm}>C_{d}^{\pm}$or $C_{d}=0$, which is addressed later.

The efficacy of the TL comparator also has a direct impact on output accuracy, especially when resolving small-valued $\Delta v_{m}$. Designing comparators with low-offset across temperature ranges and process corners is possible~\cite{Maheshwari2025} but factors, such as task complexity, $N$, mapping or TL input voltage range, can all potentially compact the computation into a smaller $\Delta v_{m}$ region around the decision boundary, pushing the comparator towards offset-generated errors.
In this paper, an \emph{instability metric}, $\Psi$, is defined as the fraction of all input samples processed that generate a $\lvert \Delta v_{m} \lvert$ less than some defined voltage tolerance level. The metric $\Psi$ can aid an IC designer in regard to required comparator accuracy and possible additional error rates. Use of a non-zero software bias, $\tau$, when training is generally useful for ACN-compatible ANNs when inputs such as $\textbf{x} = \textbf{0}$ are a valid application use case. As we shall see weight quantization during training can also be a useful tool. Integrating offset compensation into the mapping is a possibility, but offset is a function of many parameters including temperature, layout symmetry and input capacitive loading, which could vary from ACN to ACN.

Detailed error analysis of TL comparator low-offset design, impacts of capacitor quantization, SPDT switch parasitics including RC delays and the $C_{d}$ parasitic-compensation process are beyond the scope of this paper.

\subsection{Voltage swing range}
The input $PC$ signal, $V_{pc}(t)$, will ideally swing between 0V and a peak value of $V_{max}$. As such, the two membrane voltages will also swing from 0V to some modulated peak value.
Assuming no delays and an ideal comparator sampling at $t=T_{s}$ then $v_{m}$ will swing in the range
\begin{equation}
\frac{V_{max}C_{b}^{\pm}}{C_{T}^{\pm}+C_{b}^{\pm}+C_{d}^{\pm}} \le
     v_{m}^{\pm}(t=T_{s}) \le  \frac{V_{max}(C_{T}^{\pm}+C_{b}^{\pm})}{C_{T}^{\pm}+C_{b}^{\pm}+C_{d}^{\pm}}
\end{equation}

The swing in each tree will be minimal when $x_{i} = 0 \: \forall \: i$ and maximized when $x_{i} = 1 \: \forall \: i$ and dependent on $C_{b}^{\pm}$ and $C_{d}^{\pm}$. Ballast capacitor scaling has a dominant effect on the upper limit. 
When $C_{d}=0$, then the maximal swing will be $V_{max}$. The bias capacitors have a significant effect on the lower limit.


\subsection{PC voltage scaling}
\label{sec:voltage_scaling}
The single PC clock design in DTSC means computation, as defined in (\ref{eq:dtsc_proof_dtsc_hw_equation}), is  independent of $V_{pc}(t)$. This is a hugely beneficial differential property of DTSC that means $V_{pc}(t)$ magnitude can theoretically be reduced in design, or dynamically post-design, without affecting ACN output. Lower $V_{pc}(t)$ implying decreased energy consumption~\cite{Maheshwari2025}.

Real PC generator circuits are far from ideal and drops in $V_{pc}(t)$ magnitude over time and inaccurate sampling are very possible, where $V_{pc}(t=T_{s})<V_{max}$.
These non-intentional drops in PC, and consequently $\Delta v_{m}$, should  have no effect on the result.
This robustness is a significant advantage of DTSC over previous architectures like ACAN, where the use of absolute reference voltages offers no protection to these realities.
In practice, other considerations and circuit design constraints on TL, increased inaccuracies with reduced $\Delta v_{m}$, switch and supply behavior will limit PC scaling.

\subsection{Symmetry}
There is no natural symmetry in trained, real-world ANNs with $N^{+} \neq N^{-}$ and $w_{T}^{+} \neq w_{T}^{-}$ being common. However, in the conditionally-mapped DTSC ACN, some useful symmetry exists regardless.
In the mapping derivation in section~\ref{sec:conditional_mapping} it was shown 
$C_{A}^{+}=C_{A}^{-}$,
which states that the total capacitance in each tree is equal.
This symmetry is relevant for designers when performing layout of ACN capacitive arrays in custom ICs, ensuring a natural capacitive balance. Furthermore, any unwanted parasitic capacitances that are proportional to each $C_{A}^{\pm}$ will also be, in principle, easier to balance.

\subsection{Capacitive pillars}
The symmetry property of DTSC ACN trees means that fixed and equal amounts of capacitance added to the bias, or ballast, in each tree will computationally cancel out. We refer to these fixed values $C_{pb}$ and $C_{pd}$ as bias and ballast capacitive \emph{pillars}, respectively.
Capacitive pillars are a design tool that allow an IC designer to control the minimum and maximum swing of the membrane voltages without theoretically affecting computation. This is useful for ensuring the computation occurs in the voltage range best suited to the type of comparator used, which may be narrower than that of the PC.

Capacitive pillars also resolve the situation where  mapped bias and ballast capacitance values fall below the minimum allowable capacitance value, $C_{min}$, for the technology in use. Capacitance values below $C_{min}$ will not be physically realizable. For example, given mapped capacitance values of $C_{b}^{-}=30fF$ and $C_{b}^{+}=0fF$ for a MIM capacitor 180nm technology where $C_{min}=35fF$ it is possible to add a $35fF$ capacitive pillar to the bias capacitors i.e. $C_{b}^{-}=65fF$ and $C_{b}^{+}=35fF$ which are now both valid capacitance values for the technology and computation remains unaffected. The same process allows for parasitic-compensated ballast capacitors, ensuring that the compensated ballast is positive-valued.

The downside of pillars is that they reduce the dynamic range of $\Delta v_{m}$. Further, bias pillars can be used to generate an absolute voltage offset for the comparator. However, a need to generate an absolute voltage will break the voltage scaling property defined in \ref{sec:voltage_scaling}, this use should be avoided if possible, favoring comparators that are capable of working at lower differential voltages.

\subsection{Capacitive scaling and $C_{T}$ design choice}
Weight quantization in QNNs and BNNs has  reduced resources resulting in faster, lower footprint and power in digital processors~\cite{Hubara2018,Qin2020}. A primary driver for the ACN is energy-efficient  AN computation. Here AN weights are mapped to physical capacitors, which require physical IC space affecting overall chip area, device cost and energy.

The ACN capacitive voltage divider trees are  scalable. Multiplying all capacitor values in \mbox{(\ref{eq:dtsc_proof_dtsc_hw_equation})} by a constant value has no affect on computation, as the constant is  factored out.
Smaller capacitors infer smaller devices and lower device costs. It also infers less charge shuttling between PC supply and the computational array, and in principle, reduced dissipative loss.

Practical considerations, such as in integration, TL design and $C_{min}$, will limit capacitive scaling. As such,
the $C_{T}$ ACN design choice is important. Picking the same value for $C_{T}$ for all ACNs is one option. However, this is not considering the distribution of AN weight values, where small-valued weights can lead to mapped capacitance values below $C_{min}$. Trivial weight pruning is a possible solution, but destructive. Selecting each ACN $C_{T}$ such that $\min(|w_{i}|)$ maps to $C_{min}$ is another option, but this can lead to large $C_{T}$ due to large AN weight ratios i.e. $\max(|w_{i}|)/\min(|w_{i}|)$. Regularization, use of quantizers with a dead zone, or pruning during ANN weight training are alternate solutions.

\subsection{Capacitive arrays}
ACN capacitors will likely form IC capacitive arrays during chip layout, not dissimilar to those used in Successive-Approximation Register (SAR) ADC designs~\cite{Ahuja2021}. Consequently, they face similar challenges, such as device mismatch in production ~\cite{Fiorelli2016,Wakimoto2011}. Unlike SAR ADCs with regular-sized capacitance values, ACNs can generate a wide distribution of capacitances. This is an extra challenge for IC designers, with more complex layouts and routing required.

Mapped ballast capacitors are specifically problematic, as their values vary significantly with the direction of $\textbf{w}$ and in proportion to $C_{T}$. Use of ballast pillars for $v_{m}$ swing control can further increase $C_{d}$ size. Implementation of large individual capacitors can come with their own practical issues and effects during IC manufacturing and during operation.

\begin{figure}[ht]
  \centering
  \includegraphics[width=.45\textwidth]{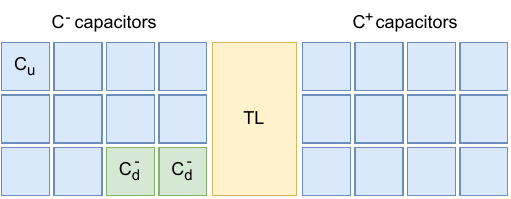}
    \caption{Illustrative $N=22$ DTSC ACN capacitive array IC layout with $N^{+}=12$ and $N^{-}=10$ with equally-sized synapse unit capacitors, $C_{u} = C_{T}/22$ and conditionally-mapped ballast capacitors $C_{d}^{-}=(N^{+}-N^{-})C_{u}$.}
    \label{fig:dtsc_layout_bnn22}
\end{figure}

Building arrays using tiled, composite, equally-sized capacitors is one design approach~\cite{Chen2017}. If quantization during ANN training is used, more regularly-shaped arrays can be formed with less quantization errors. A mapped BNN with  $w_{i} \in \{-1,+1\}$, is illustrated in Fig.~\ref{fig:dtsc_layout_bnn22}. The symmetry property of a DTSC ACN is clearly shown.
The downside of increasing levels of quantization during training is that potentially more ANs will be required to achieve the same level of functionality, compared to ANNs with real-valued weights.

\section{Software}\label{sec:software}
To verify various ACN mappings, 10,000 $N=8$ real-valued weight vectors were randomly selected using a $\mathcal{N}(0,0.01)$ normal distribution; allowing for small $w_{T}$ required for the ReLU mapping. Each vector was mapped and then tested with all possible 256 inputs with any errors between AN and ACN outputs recorded. Table ~\ref{tab:rand-weight-mapping-results} provides results for 7 different mapping configurations. All software weights mapped correctly across all configurations, except for 12 errors apportioned to  numeric rounding. The \emph{ReLU} mapping had 22 weight vectors with excessive  $w_{T}$ that failed to map. As predicted, conditional mapping provided significantly better results than other options with minimal $C_{d}=\sum C_{d}^{\pm}$, maximal $\lvert \textbf{C} \lvert$ and, consequently, $\Delta v_{m}$. Capacitive pillars were shown to work, with obvious extra capacitive cost and lower $\lvert \textbf{C} \lvert$.
\begin{table}[t]
\caption{10,0000 Random $\mathcal{N}(0,0.01)$ $N=8$ weight vector mapping testing with $C_{T}=100fF$, and capacitive pillars (if used) $C_{pb}=2fF$,$C_{pd}=5fF$, over all possible 256 input patterns \label{tab:rand-weight-mapping-results}}
\centering
\begin{tabular}{|p{0.22\textwidth}|p{0.07\textwidth}|p{0.07\textwidth}|}
\hline
Mapping configuration & Mean $C_{d}$ & Mean $\lvert \textbf{C} \lvert$\\
 & (Dev.), $fF$ & (Dev.), $fF$\\
\hline \hline
Conditional,  $\tau=0.0$ &  36 (25) & 0.66 (0.12) \\ \hline
Conditional vectored-bias,  $\tau=0.1$  & 32 (23) & 0.62 (0.11) \\ \hline
Conditional, $\tau=0.1$ & 52 (26) & 0.56 (0.08) \\ \hline
Conditional, $\tau=0.1$, and pillars & 62 (26) & 0.52 (0.07) \\ \hline
Conditional, $\tau=-0.1$ & 52 (26) & 0.56 (0.08) \\ \hline
Balanced, $\tau=0.1$ & 117 (5) & 0.40 (0.03) \\ \hline
ReLU, $\tau=0.1, V_{max}=1.0V$ & 187 (72) & 0.29 (0.06) \\ \hline
\end{tabular}
\end{table}

 Results for conditional mapping and varying $N$ are given in Table~\ref{tab:rand-weight-mapping-results-bnn}. In this synthetic test, real-valued weights were also binarized to $\{-1,1\}$ for comparison. Results show that the average $\lvert \textbf{C} \lvert$ decreased with increasing $N$, as expected. In this case binary weights, as predicted, has also led to smaller $\lvert \textbf{C} \lvert$, even though they required smaller $C_{d}$. Mean $C_{d}$ also reduces with increasing $N$. However, weights were drawn from a \emph{symmetric}  distribution, where \mbox{$mean \, \lvert w_{T}^{+}-w_{T}^{-} \lvert/w_{T}$} will naturally tend towards zero with increasing $N$. 
\begin{table}[t]
\caption{10,0000 Random $\mathcal{N}(0,0.01)$ conditional weight vector mapping ($\tau = 0.0$) testing for different $N$ with $C_{T}=100fF$ against equivalent binarized $\{-1,1\}$ version \label{tab:rand-weight-mapping-results-bnn}}
\centering
\begin{tabular}{|p{0.03\textwidth}|p{0.07\textwidth}|p{0.07\textwidth}|p{0.07\textwidth}|p{0.07\textwidth}|}
\hline
    & \multicolumn{2}{c|}{Real-valued weights} & \multicolumn{2}{c|}{Binarized weights} \\ \cline{2-5}
$N$ & Mean $C_{d}$ & Mean $\lvert \textbf{C} \lvert$ & Mean $C_{d}$ & Mean $\lvert \textbf{C} \lvert$\\
& (Dev.), $fF$ & (Dev.), $fF$ & (Dev.), $fF$ & (Dev.), $fF$\\
\hline \hline
8 & 36 (25) & 0.66 (0.12) & 27 (22) & 0.57 (0.10) \\ \hline
16 & 25 (18) & 0.50 (0.07) & 20 (15) & 0.42 (0.05) \\ \hline
32 & 18 (13) & 0.38 (0.04) & 14 (11) & 0.31 (0.03) \\ \hline
64 & 13 (9) & 0.28 (0.02) & 10 (8) & 0.23 (0.01) \\ \hline
784 & 4 (3) & 0.09 (0.00) & 3 (2) & 0.07 (0.00) \\ \hline
\end{tabular}
\end{table}

Consequently, we explore the conditional-mapping behavior when using  trained ANN weights that embed information in their distributions.
Three different 2-layer ANN configurations were trained in Python 3.11.2 using a combination of TensorFlow (version 2.12.0) and the Larq (version 0.13.0) extension for QNNs and BNNs. 
Networks were trained on a binarized version of the  28x28 pixel MNIST image classification dataset. The MNIST binarization used binary '1' to represent pixel values above 127, binary '0' otherwise. The networks were trained with an Adam optimizer for 50 epochs using 60,000 training samples and 10,000 validation samples. Each network was trained five times with different initialized weights.

All trained ANNs had $N=784$ \mbox{1-bit} inputs and 10 linear output neurons with real-valued weights. ANs in the hidden layer used an ACN-compatible binary activation function and weights quantized differently per configuration. This demonstrated support for different ANNs, but also the impact of quantization on ACN properties, such as overall chip size and accuracy.
The number of hidden layer ANs was selected to achieve similar classification performance across configurations.
ANN configurations are provided in Table~\ref{tab:mnist-nn-config}.
\begin{table}[t]
\caption{MNIST Experimental Network Configurations With ACN-Compatible 784-bit Input/1-bit Output Hidden Layer ANs\label{tab:mnist-nn-config}}
\centering
\begin{tabular}{|c|p{7cm}|}
\hline
Name & Network configuration \\
\hline \hline
part32 & ANN with 32 hidden-layer neurons with real-valued weights and biases. \\ \hline
kbit36 & QNN with 36 hidden-layer neurons with real-valued bias and 4-bit quantized weights with a [1,-1] clipping constraint. \\ \hline
bin156 & BNN with 156 hidden-layer neurons with binary \mbox{\{-1,+1\}} weights and fixed bias, $\tau = 0.5$.\\ \hline
\end{tabular}
\end{table}
QNN/BNN networks \emph{kbit36} and \emph{bin156} used the Larq quantizers \textbf{DoReFaQuantizer} and \textbf{SteSign} respectively~\cite{Geiger2020}.

All hidden AN weights and biases were extracted from the trained models and conditionally mapped with constraints \mbox{$C_{min}=2fF$} and $V_{max}=1.0V$. To ensure mapped ballast and bias capacitors are realizable ($\ge C_{min}$), bias and ballast capacitive pillars of $2fF$ were used.

Mapping \emph{part32} ANNs created an issue due to some weights being very small-valued, implying a very large $C_{T}$ to ensure all $C_{i}$ were realizable. Therefore, trivial weight pruning of $\lvert w_{i} \lvert$ below a threshold was used. This meant finding a balance between loss of classification performance and minimizing $C_{T}$. A $C_{T}=4704fF$ was chosen with a pruning threshold of 0.078, which led to an average $1\%$ drop in overall classification but resulted in all ACN capacitors being realizable. The pruning process removed, on average, 44\% of the weights.

The advantages of ANN quantization are now obvious. Quantization removes extremely small-valued and unrealizable capacitance values. The minimum $\lvert w_{i} \lvert$ is now determined by the quantizer. Setting \mbox{$C_{T} = C_{min} w_{T}/min(|w_{i}|)$} for each $N$ assures all $C_{i}$ will be realizable. For \emph{kbit32} ANNs this resulted in an average \mbox{$C_{T}=4976fF$} with values ranging from $2198fF$ up to $6597fF$. The network now has ACNs with significant variations in size. These irregularly-sized ACNs are now adding a new IC layout challenge. Note, bias values are not quantized in Larq \textbf{QuantDense} layers, which can lead to very small-valued trained biases. As such, using a vectored-bias mapping resulted in $C_{T}$ values up to $35,240fF$.

For the \emph{bin156} ANNs $C_{i}^{\pm} = C_{min}$ and $C_{T}=1568fF$. The bias capacitor is calculated as $C_{b}^{-} = \tau C_{min} = 1fF$. This "half-C" bias prevents $\Delta v_{m} = 0$ regardless of input. As $C_{b}^{-} < C_{min}$, the $2fF$ bias and ballast pillar applied to each tree maintains this functionality.

Binary-MNIST classification accuracy results are shown in Table~\ref{tab:mnist-nn-results}. The TensorFlow  ANN classification over 10,000 images was compared against the ANN inference when the hidden layer was replaced by the ACN mapped hardware model. The 1\% drop in accuracy with \emph{part32} was expected due to pruning.
\begin{table}[t]
\caption{MNIST Results With Validation Classification Accuracy For Both Original Software And Mapped ACN Hardware Model In Hidden Layer and $5mV$ Instability Metric $\Psi$\label{tab:mnist-nn-results}}
\centering
\begin{tabular}{|c|c|c|c|}
\hline
Name & Mean Software & Mean Hardware & Instability \\
     & (Dev.) (\%) &  (Dev.) (\%) & Metric, $\Psi$ \\
\hline \hline
part32 & 92.0 (0.2) & 91.0 (0.2) & 0.08 \\ \hline
kbit36 & 92.6 (0.2) & 92.6 (0.2) & 0.02 \\ \hline
bin156 & 92.1 (0.4) & 92.1 (0.4) & 0.01 \\ \hline
\end{tabular}
\end{table}
The instability metric shows for the \emph{part32} ANN nearly 8\% of computations would require the physical comparator circuit to resolve $\Delta v_{m}$ below $5mV$. Quantization has significantly reduced instability, meaning fewer computations are potentially affected by comparator inaccuracies.

Table~\ref{tab:mnist-capvalue-results} shows the capacitance values for the mapped ACNs.
The mean size of the ACN hidden layer increased with greater quantization and hidden neurons. 
However, \emph{bin156} was only roughly 50\% greater than \emph{part32} due to the smaller $C_{T}$ used, even though there were nearly 5 times the number of ANs.
\begin{table}[t]
\caption{MNIST Hidden Layer Mapped ACN Capacitance Value Sizes \label{tab:mnist-capvalue-results}}
\centering
\begin{tabular}{|c|c|c|c|c|}
\hline
Name & Mean $C_{T}$ & Mean $C_{d}$ & Mean $\lvert \textbf{C} \lvert$ & Mean Layer  \\
     & (Dev.), $fF$ & (Dev.), $fF$ & (Dev.), $fF$ & Size (Dev.), $pF$ \\
\hline \hline
part32 & 4704 (0) & 1338 (101) & 0.10 (0.00) & 194 (3) \\ \hline
kbit36 & 4976 (114) & 1718 (48) & 0.08 (0.00) & 242 (5) \\ \hline
bin156 & 1568 (0) & 367 (8) & 0.06 (0.00) & 303 (1) \\ \hline
\end{tabular}
\end{table}

\begin{figure}[ht]
  \centering
  \includegraphics[width=.4\textwidth]{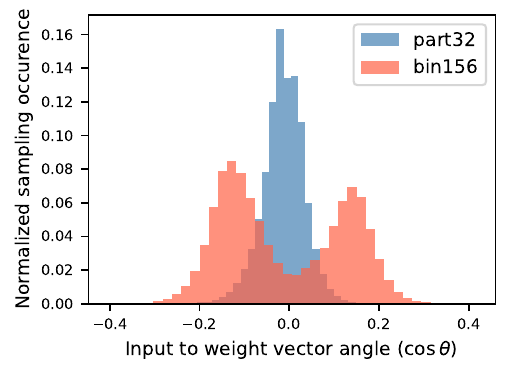}
    \caption{Sampling distribution of $\cos \theta$ over all 10,000 validation samples for both the \emph{part32} and \emph{bin156} networks.}
    \label{fig:cost_hist_part_vs_bnn}
\end{figure}

Table~\ref{tab:mnist-dvm-results} gives $\lvert \Delta v_{m} \lvert$ statistics over all the test images and ACNs. As  predicted by 
$\lvert \textbf{C} \lvert$ in Table~\ref{tab:mnist-capvalue-results}, the maximum $\lvert \Delta v_{m} \lvert$ reduces with increased levels of quantization. However, the average \emph{part32} $\lvert \Delta v_{m} \lvert$ is half that of the quantized ANNs. From \mbox{(\ref{eq:dvm_as_expanded_dot_product})} this infers a difference in $\cos\theta$, as the inputs are common between tests. Fig.~\ref{fig:cost_hist_part_vs_bnn} shows the distribution of $\cos\theta$ over all tests for the \emph{part32} and \emph{bin156} networks. The direction of trained ANN weight vectors again has a significant effect on the characteristics of physical ACNs. Increased correlation of weight vectors with inputs in \emph{bin156} with the larger number of ANs is reducing potential comparator inaccuracies with, on average, larger $\lvert \cos\theta \lvert$ and $\lvert \Delta v_{m} \lvert$.
\begin{table}[t]
\caption{MNIST Absolute Differential Membrane Voltages Over 10,000 Validation Images\label{tab:mnist-dvm-results}}
\centering
\begin{tabular}{|c|c|c|}
\hline
Name & Mean $\lvert \Delta v_{m} \lvert$ & Max $\lvert \Delta v_{m} \lvert$ \\
     & (Dev.) ($mV$) & ($mV$) \\
\hline \hline
part32 & 39 (0.04) & 481 \\ \hline
kbit36 & 75 (0.06) & 431 \\ \hline
bin156 & 75 (0.04) & 337 \\ \hline
\end{tabular}
\end{table}

\section{Discussion}\label{sec:discussion}
The energy-efficient DTSC ACN uses an SC network to perform computation using charge provided by a time-varying PC supply. From the PC perspective, the ACN network can be modeled as an RC load circuit comprising of a dynamic parasitic resistive load ($R_{L}$) and a dynamic switched capacitive load ($C_{L}$). The dissipative energy, and consequently PC efficiency, is a function including these two variables~\cite{Maheshwari2025}. The $C_{L}$ will be a function of input but also mapping design choices such as $C_{T}$, the mapping algorithm and  ANN weight vectors. Minimizing the ACN capacitance values, through an optimal mapping, is not only important for reducing both the chip area size and potential frequency fluctuations in the PC supply but also energy consumption.

The paper has provided a rigorous mathematical introduction for an IC designer on mapping ANs to ACNs, as well as practical tools and pointers. DTSC ACN requires only $N$ synapse capacitors compared to $2N$ solutions~\cite{Massarotto2024} and the mapping provides an exact value for $C_{d}$. We have shown how metrics, such as $\Psi$ and $\lvert \Delta v_{m} \lvert$ expose the demands on TL performance. The introduction of SPDT switches in DTSC also provides a linear relationship between $\Delta v_{m}$ and the input, like the ANN $\textbf{w}\cdot\textbf{x}$ dot product.
Finally, the ACN allows IC designers to build arbitrary-sized capacitors to represent arbitrary-precision ANN weights. However, as we have seen, there are also significant advantages to weight/capacitance value quantization. In the extreme, binary weighted, regularly-shaped, ACNs will look attractive to IC designers familiar with ADC arrays. Smaller, binarized ACN designs, crucially reduce capacitor design burden but will need to consider the energy trade-off
with an increase in the number of ACNs required to achieve equivalent levels of functionality.

Going forward, thoughts turn to how large, multiple ACN, multi-layer networks will be physically constructed and their associated cost. Mismatch IC issues may be problematic, as well as input routing, PC distribution and ability to accurately instantiate required capacitance values. Networks will need to tackle increasingly demanding applications. More generic, reusable, weight configurable and re-programmable ACNs will be required. Thought should also be given to more ACN-friendly software training tools. Given knowledge of the mapping properties, thought can now be applied to ACN-specific tools, such as quantizers, regularizers and pruning techniques, that promote smaller values for $C_{T}$, or punish weights that unnecessarily promote large mapped ballast capacitance values or small $\Delta v_{m}$.
Consideration will also be required with respect to more complex networks, such as the ubiquitous Convolutional Neural Network (CNN). Support for multi-bit output ACNs and popular activation functions, such as ReLU, will need to be investigated in a drive towards increasingly efficient and cost effective computation. The present work, thus, lays a key foundation stone in a wider field of study where we see significant potential for innovation.

\section{Conclusion}\label{sec:conclusion}
This paper has demonstrated that ANNs trained using standard tools, such as TensorFlow, with real-valued, quantized or binary signed weights, can be mapped optimally and robustly onto energy-efficient, adiabatic, DTSC ACNs. 
Specifically, we have shown that choosing DTSC ACNs as the fundamental adiabatic neuron unit implementation and combining it with the conditional mapping approach we can minimize total capacitance and chip area, 
and maximize the differential voltages applied to physical, non-ideal comparators.
We have seen how weight quantization can play a significant role in generating practical and realizable designs.
Lastly, 
properties like symmetry, voltage and capacitive scaling,  as well as tools such as capacitive pillars, make DTSC ACNs an attractive solution for designers looking for energy-efficient solutions.

Although challenges remain, the future of adiabatic logic and charge recovery appears ever more promising for the design and implementation of large ANNs, as well as other computationally intensive applications. Furthermore, the regularity and algorithmic nature of the mapping process holds tantalizing prospects for future automation of the design of DTSC ACN networks.

\bibliographystyle{IEEEtran}
\bibliography{export}
\newpage
\section{Biography Section}

\vskip -3\baselineskip plus -1fil

\begin{IEEEbiography}
[{\includegraphics[width=1in,height=1.25in,clip,keepaspectratio]{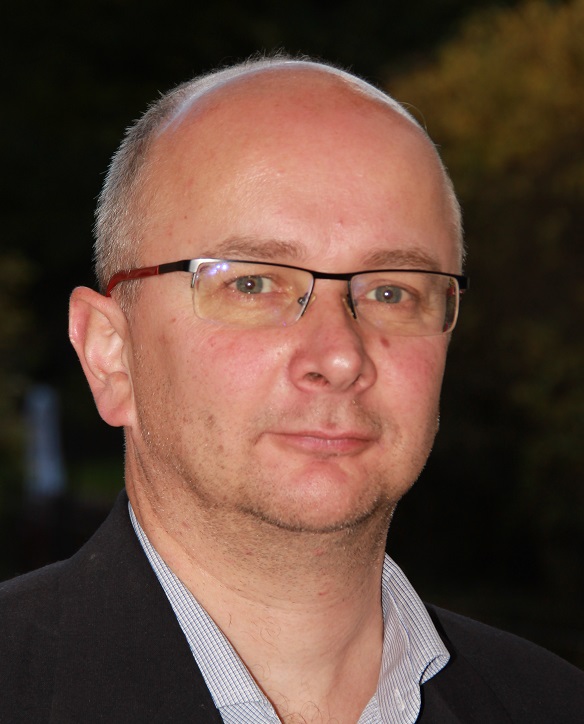}}]{Mike Smart} received the degree of electrical engineering from the University of
Warwick, U.K. and the Ph.D. degree in electrical engineering from the University
of Edinburgh, U.K., in 1992 and 1996 respectively. He has worked as a staff engineer for Motorola Solutions and a lead engineer for IndigoVision Ltd.
He is currently a Software Engineer at the University of Edinburgh, U.K. His research
interests include AI, novel computation, algorithms, video
compression and biologically-inspired software.
\end{IEEEbiography}

\vskip -2\baselineskip plus -1fil
\begin{IEEEbiography}
[{\includegraphics[width=1in,height=1.255in,clip,keepaspectratio]{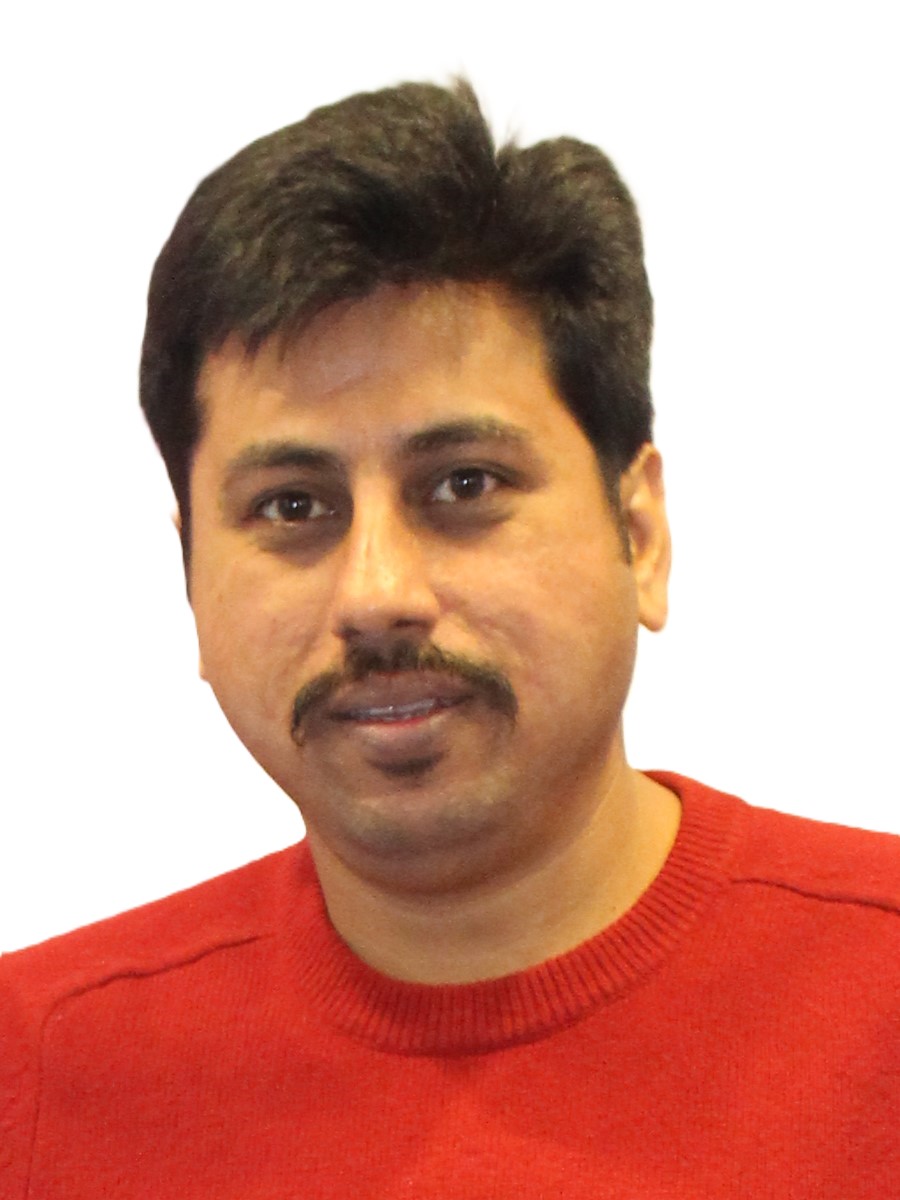}}]{Sachin Maheshwari}
(S’12–M’21) received a B.E. in Electrical and Electronic Engineering from ICFAI University, an M.E. in Microelectronics from BITS Pilani, and a Ph.D. in Electronics Engineering from the University of Westminster, UK. He was a Research Fellow at the University of Southampton and is currently a Research Associate at the Centre for Electronics Frontiers, University of Edinburgh. His research focuses on neuromorphic computing and artificial neural networks, with emphasis on energy-recovery logic (adiabatic techniques) and emerging technologies (RRAM) for energy-efficient brain-inspired systems. 
\end{IEEEbiography}

\vskip -2\baselineskip plus -1fil

\begin{IEEEbiography}
[{\includegraphics[width=1in,height=1.25in,clip,keepaspectratio]{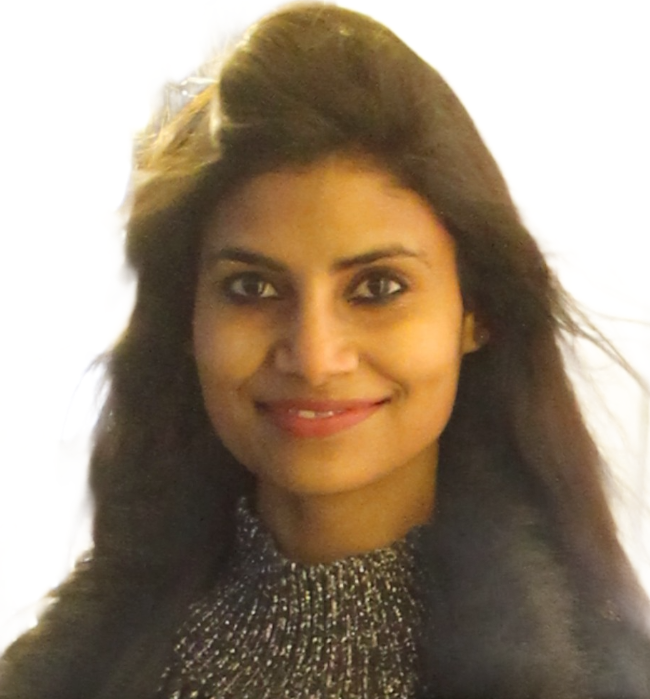}}]{Himadri Singh Raghav} received her B.Sc. and M.Sc. in electronics and M.Tech. in VLSI Design from Banasthali University, Rajasthan, India. She then obtained her Ph.D. in Electronics Engineering from the University of Westminster, London, UK. She worked for 3 years as a Research Fellow at the National University of Singapore. She is currently working as a Research Associate with the Centre for Electronics
Frontiers, School of Engineering, University of Edinburgh, UK. Her research interest is in energy efficient implementation of a secure system using charge recovery logic
\end{IEEEbiography}

\vskip -2\baselineskip plus -1fil

\begin{IEEEbiography}
[{\includegraphics[width=1in,height=1.25in,clip,keepaspectratio]{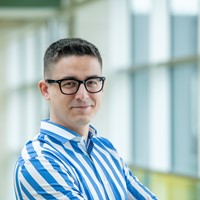}}]{Alexander Serb}
received his degree in Biomedical Engineering from Imperial College in 2009 and his PhD in Electrical and Electronics Engineering from Imperial College in 2013. Currently, he is a reader at the University of Edinburgh, UK. His research interests are: cognitive computing, neuro-inspired engineering, algorithms and applications using RRAM, RRAM device modelling and instrumentation design.
\end{IEEEbiography}

\vfill

\end{document}